\documentstyle[preprint,eqsecnum,aps]{revtex}

\begin{document}
\draft

\title{ON THE TWO- AND THREE DIMENSIONAL LENZ-ISING-ONSAGER PROBLEM IN
PRESENCE OF MAGNETIC FIELD }

\author{Martin S. Kochma\'nski\\}
\address{Institute of Physics, 
University of Rzesz\'ow\\
T.Rejtana 16 A, 35--310 Rzesz\'ow, Poland\\
e-mail: mkochma@atena.univ.rzeszow.pl}
\maketitle

\begin{abstract}
In this paper a new  approach to solving the Ising-Onsager problem  in external
magnetic field   is investigated. The expression for free energy on one Ising
spin in external field both for the twodimensional and threedimensional Ising
model with interaction  of the nearest neighbors are derived. The
representations of free energy being expressed by multidimensional  integrals of
Gauss type with the appropriate  dimensionality are shown. Possibility of
calculating the integrals and the critical indices on the base of the derived
representations for free energy is investigated.
\end{abstract}
\widetext
\begin{center}
INTRODUCTION
\end{center}
It is well known that the Ising-Onsager problem          $\cite{ising25},
\cite{onsager44}$
in external magnetic field  has not been solved till now, despite intensive
efforts  of a few generations of physicists and mathematicians. By the problem
we mean the exact calculation of the statistical sum  for the Ising model with
interaction of the nearest neighbors both in the twodimensional case (in
finite external field) and in the threedimensional case  (in external field 
as well as without such a field). Therefore, we do not intend to present the 
variety of
approximate methods  and approaches to solving the Ising-Onsager problem.
Detailed discussion of these  matters could be found in numerous well known
papers and monographs. We note here only the paper by C.N. Yang $\cite 
{yang52}$, where the
problem was investigated for the case of infinitely small external field in 
two dimensions. In our opinion, the efforts to find an exact solution to the
Ising-Onsager problem is of great interest till now. The reason is, the Ising
models are connected with numerous other models of statistical mechanics and
quantum field theory
           $\cite{gaudin83}-\cite{thompson88}$.
Therefore  the present paper we treat as the next step in the direction
described above.
\widetext
\section{THE PARTITION FUNCTION}
\label{sec:level1}
\widetext
Here we consider firstly the twodimensional case
$(d=2, H\neq 0)$, and  then we will report on the results for 
the threedimensional case $(d=3, H\neq 0)$
without going into details of derivations. Let us consider  a rectangular
lattice consisting of $M$ columns and $N$ lines,
in nodes of which are given variables $\sigma_{nm}$ which take values
$\pm 1$.

These variables will be called below "spins". The collective index $nm$ numbers
nodes of the lattice; $n$ numbers a line, $m$ numbers a column. The Ising model
with nearest   neighbors interaction is given by the following form of the
Hamiltonian:
\begin{equation}
{\cal H}=-J_{2}\sum_{nm}\sigma_{nm}\sigma_{n+1,m}-J_{1}\sum_{nm}
\sigma_{nm}\sigma_{n,m+1}-H\sum_{nm}\sigma_{nm},
\end{equation}
which takes into account possible anisotropy of the interaction between nearest
neighbors and also interaction of spins   $\sigma_{nm}$   with external  field
$H$, which is directed "upwards" $(\sigma_{nm}=+1)$.
The investigated problem consists of calculation of the statistical sum for 
the system:
\begin{eqnarray*}
Z(h)=\sum_{\sigma_{11}=\pm 1}...\sum_{\sigma_{NM}=\pm 1}\exp\left(-
\beta{\cal H}\right)
\end{eqnarray*}
\begin{equation}
=\sum_{(\sigma_{nm}=\pm 1)}\exp\left[\sum^{NM}_{n,m=1}(K_{2}\sigma_{nm}
\sigma_{n+1,m}+K_{1}\sigma_{nm}\sigma_{n,m+1}+h\sigma_{nm})\right],
\end{equation}
where
\begin{equation}
K_{1,2}={\beta}J_{1,2}, \;\;\;\;\; h={\beta}H, \;\;\;\; \beta=1/k_{B}T .
\end{equation}
Typically periodic boundary conditions on variables
 $\sigma_{nm}$
 are imposed and we will assume this everywhere below.
 Let us note that the statistical sum $(1.2)$ is symmetric with respect to the
 change
 $h\rightarrow -h$.

As is known $\cite{sml64}$, the statistical sum $(1.2)$
can be represented in the form of the trace of the
                             $T$-operator ($T$-{\it transfer matrix}):
\begin{equation}
Z(h)=Tr(T)^{M}=Tr\left[(2\sinh 2K_{1})^{(N/2)}T_{1}T_{2}T_{h}\right]^{M},
\end{equation}
where the matrices $T_{1,2,h}$ are of the form:
\begin{equation}
T_{1}=\exp\left(K^{*}_{1}\sum^{N}_{n=1}\sigma^{x}_{n}\right),
\end{equation}
\begin{equation}
T_{2}=\exp\left(K_{2}\sum^{N}_{n=1}\sigma^{z}_{n}\sigma^{z}_{n+1}\right), \;\;
\;\;\;\; \sigma^{z}_{N+1}=\sigma^{z}_{1},
\end{equation}
\begin{equation}
T_{h}=\exp\left(h\sum^{N}_{n=1}\sigma^{z}_{n}\right),
\end{equation}
and $K^{*}_{1}$ and $K_{1}$
are connected by the relations:
\begin{equation}
 \exp(-2K_{1})=\tanh(K^{*}_{1}), \;\;\;\; \sinh(2K_{1})\sinh(2K^{*}_{1})=1 .
\end{equation}
In the formulae $(1.5)-(1.7)$ the quantities $(\sigma^{x,y,z}_{n})$, $(n=1,2,...N)$
- are well known from quantum mechanics $2^{N}$-dimensional matrices:
\begin{eqnarray*}
\sigma^{x,y,z}_{n}=1\otimes 1\otimes ...\otimes \sigma^{x,y,z}
\otimes ... \otimes 1, \;\;\;\; (N \;\;{\rm factors}),
\end{eqnarray*}
where $\sigma^{x,y,z}$--- twodimensional spin  Pauli  matrices:
\begin{equation}
\sigma^{x}=\left(\begin{array}{cc}0 & 1\\ 1 & 0\end{array}\right), \;\;\;\;
\sigma^{y}=\left(\begin{array}{cc}0 & -i\\ i & 0\end{array}\right),
\;\;\;\;
\sigma^{z}=\left(\begin{array}{cc}1 & 0\\ 0 & -1\end{array}\right),
\end{equation}
satisfying the standard transposition relations:
\begin{equation}
(\sigma^{k})^2=1, \;\;\; \sigma^{k}\sigma^{j}+\sigma^{j}\sigma^{k}=0,
\;\;\;\; (k,j)=x,y,z; \;\;\;;\ \sigma^{x}\sigma^{y}=i\sigma^{z},\; ... 
\end{equation}
For example, spin matrices for the $n$-th electron in  the system consisting of
$N$ nonrelativisitc electrons are exactly the matrices  $\sigma^{x,y,z}_{n}$.
It is known that for
 $n\neq n^{'}$ spin matrices $\sigma^{x,y,z}_{n}$
commute and for any given particular
 $n$ they satisfy formally relations
 $(1.10)$. It follows from this that matrices
 $T_2$ and $T_h$, $(1.6-7)$
 commute but they do not commute with the matrix
 $T_1$, $(1.5)$, i.e.
\begin{equation}
[T_2,T_h]_{-}=0, \;\;\;\; [T_h,T_1]_{-}\neq 0, \;\;\;\; [T_2,T_1]_{-}\neq 0 .
\end{equation}

It follows from $(1.11)$ that under $Tr(...)$
we can write for the statistical sum
 $(1.4)$ the expression:
\begin{equation}
Z(h)=Tr(\gamma T_{1}T^{1/2}_{h}T_{2}T^{1/2}_{h})^{M}=
Tr(\gamma T^{1/2}_{h}T_{1}T^{1/2}_{h}T_{2})^{M}\equiv Tr(P)^{M},
\end{equation}
\begin{equation}
P\equiv (2\sinh 2K_{1})^{N/2}T^{1/2}_{h}T_{1}T^{1/2}_{h}T_{2},
\end{equation}
where we used the identity $Tr(AB)\equiv Tr(BA)$.

Now we will consider in more details the matrix $U\equiv
T^{1/2}_{h}T_{1}T^{1/2}_{h}$;
since the matrices
 $\sigma^{x}_{n}$ and $\sigma^{z}_{n^{'}}$,
 commute for $n\neq n^{'}$,
  we can write the matrix $U$  in the form:
\begin{equation}
U\equiv T^{1/2}_{h}T_{1}T^{1/2}_{h}=\prod^{N}_{n=1}e^{(h/2)\sigma^{z}_{n}}
e^{K^{*}_{1}\sigma^{x}_{n}}e^{(h/2)\sigma^{z}_{n}}\equiv\prod^{N}_{n=1}U_{n}.
\end{equation}
Further the matrix
 $U_n$ can be represented in the following form:
\begin{equation}
U_n=e^{(h/2)\sigma^{z}_{n}}(\cosh K^{*}_{1}+\sigma^{x}_{n}\sinh K^{*}_{1})
e^{(h/2)\sigma^{z}_{n}}=\exp\left[\omega\left(\frac{\cosh K^{*}_{1}\sinh h}
{\sinh\omega}\sigma^{z}_{n}+\frac{\sinh K^{*}_{1}}{\sinh\omega}\sigma^{x}_{n}
\right)\right],
\end{equation}
where $\omega$ -  is a positive root of the equation:
\begin{equation}
\cosh\omega=\cosh K^{*}_{1}\cosh h,
\end{equation}
In determination of the expression $(1.15)$ we used the identity:
\begin{equation}
\exp(\mu t)=\cosh\mu+t\sinh\mu, \;\;\;\; t^2=1.
\end{equation}
It is easy to see that for
 $h=0$ we obtain from $(1.12)$
 the standard expression for the  statistical sum
$Z$ for the twodimensional Ising model without external field
$\cite{baxter82,izyum87}$.
\subsection{The $1D$ Ising model}

Relatively easy one can show that $(1.12)$ becomes:
\begin{equation}
Z(h)=Tr\left[(2\sinh 2K_{1})^{N/2}\prod^{N}_{n=1}U_{n}T_{2}\right]^{M},
\end{equation}
where $U_{n}$ is given by the formula $(1.15)$,
and it describes correctly the transition to the onedimensional Ising model
both in the constant $K_{1}$, and in the constant $K_{2}$.
Indeed, if we take   $K_{2}=0$ and $N=1$,
and this procedure corresponds to  neglecting summation over $n$ we obtain:
\begin{equation}
Z_{1}(h)=Tr\left[(2\sinh 2K_{1})^{1/2}U_{0}\right]^{M},
\end{equation}
where the matrix $U_{0}$ is given by  $(1.15)$,
where the index $n$ was omitted. Eigenvalues of the matrix $U_0$ can be easily
obtained:
\begin{eqnarray*}
\lambda^{\pm}=\exp(\pm\omega),
\end{eqnarray*}
where $\omega$ --- is the positive root for the equation $(1.16)$.
As a result we obtain the following formula describing free energy on one 
spin in the thermodynamic limit:
\begin{equation}
f(h)=-\frac{1}{\beta}\lim_{M\rightarrow\infty}\frac{1}{M}\ln Z_{1}(h)=
-\frac{1}{\beta}\ln\left[e^{K_{1}}\cosh h +(e^{2K_{1}}\sinh^{2}h+e^{-2K_{1}})^
{1/2}\right],
\end{equation}
i.e.  the known classical expression $\cite{ising25}$.

Transition to the onedimensional Ising model limit in the constant $K_1$ could
be done by taking
$K_{1}\to 0$, and $M=1$, i.e.
we neglect summation over the index $m$ and go to the limit $K_{1}\to 0$.
As a result we get from  $(1.18)$:
\begin{equation}
Z_{1}(h)=Tr\left[\prod^{N}_{n=1}(1+\sigma^{x}_{n})T_{2}T_{h}\right],
\end{equation}
where we used the transition to the limit:
\begin{eqnarray*}
\lim_{K_{1}\rightarrow 0}(2\sinh 2K_{1})^{1/2}\exp(K^{*}_{1}\sigma^{x}_{n})
=(1+\sigma^{x}_{n}),
\end{eqnarray*}
where we took into account the relation
 $(1.8)$ between $K_{1}$ and $K^{*}_{1}$.
 It is reasonable  to stress here that the factors
 $(1+\sigma^{x}_{n})$,
  entering the expression
  $(1.21)$, are simply necessary to get the correct result. For the aim of
  calculating the trace
  $(1.21)$,
  it is convenient to go to the fermion representation
$\cite{izyum87,sml64}$. 
Omitting some calculations we write for $Z_{1}(h)$, $(1.21)$
the expression:

\begin{equation}
Z_{1}(h)=Tr(DT^{\pm}_{2}T_{h}),
\end{equation}
where the operators $D$, $T^{\pm}_{2}$ and $T_{h}$,
 expressed in terms  of Fermi creation and annihilation  operators
 $(c^{+}_{n},c_{n})$
are of the form:
\begin{equation}
D=\prod_{n=1}^{N}\left[1+(-1)^{c_{n}^{+}c_{n}}\right].
\end{equation}
\begin{equation}
T_{2}^{\pm}=\exp\left[K_{2}\sum_{n=1}^{N}(c_{n}^{+}-c_{n})(c_{n+1}^{+}+
c_{n+1})\right],
\end{equation}
\begin{eqnarray}
T_{h}=\exp\left\{h\sum_{n=1}^{N}\exp\left[i\pi\sum_{p=1}^{n-1}
c_{p}^{+}c_{p}\right]
(c_{n}^{+}+c_{n})\right\},
\end{eqnarray}
In the formula $(1.24)$ the sign $(+)$
is related to states  that are even with respect to the operator of the complete
number of particles  $(\hat{N}=\sum^{N}_{n=1}c^{+}_{n}c_{n})$,
to which there correspond anticyclic boundary conditions, and the sign $(-)$
to the odd states, to which there correspond cyclic boundary conditions.
It is easy to see that because of the multiplicative character of the operator
                              $D$, $(1.23)$,
 all diagonal matrix elements in  $(1.22)$
vanish with the exception of the vacuum-vacuum matrix element, i.e.:
\begin{equation}
Z_{1}(h)=2^{N}<0\mid(T_{2}^{\pm}T_{h})\mid0>,
\end{equation}
where the operators $T^{\pm}_{2}$ and $T_{h}$
are defined by $(1.24-25)$.
Then, "acting" with the operator  $T_{h}$ on the vacuum state $\mid0>$,
and using the Hausdorff-Baker formula $\cite{glimm81}$,
(${\alpha},{\beta}={\it const}$):
\begin{eqnarray*}
\exp({\alpha}x)\exp({\beta}y)=\exp({\alpha}x+{\beta}y+({\alpha}{\beta}/2)
[x,y]_{-}),
\end{eqnarray*}
\begin{eqnarray*}
[x,z]_{-}=[y,z]_{-}=0, \;\;\;\;\;\; z\equiv[x,y]_{-},
\end{eqnarray*}
the operator $T_{h}$, $(1.25)$
can be reduced to the "effective" form (in the sense of action on
 $\mid0>$):
\begin{equation}
T_{h}=\cosh^{N}(h)\exp\left[\tanh^{2}h\sum_{n=1}^{N}\sum_{p=1}^{N-n}
c_{n}^{+}c_{n+p}^{+}\right].
\end{equation}
When developing the expression $(1.27)$ we have taken into accaunt the fact, 
that the diagonal matrix elements of the odd number of Fermi operators are 
equal to zero. 

Finally, going to the momentum representation: 
\begin{eqnarray*}
c_{n}=\frac{\exp(-i\pi/4)}{\sqrt{N}}\sum_{q}e^{{i}q\*n}
{\eta}_{q},
\end{eqnarray*}
and computing the matrix element for  a fixed $q$ after some not complicated
transformations we arrive in the case of even states at  the expression
for the statistical sum $[Z^{+}_{1}(h)]$  $(1.26)$:
\begin{eqnarray}
Z^{+}_{1}(h)=[2\cosh(h)]^{N}{\prod_{0{\leq}q{\leq}{\pi}}^{}}
[\cosh 2K_{2}-\sinh 2K_{2}\cos q+{\alpha}^{2}\sinh 2K_{2}(1+\cos q)]\nonumber\\
=[2{\cosh(h)}{\cosh K_{2}}]^{N}\prod_{n=1}^{N}\left[1+z_{2}^{2}+2z_{2}z-
2z_{2}(1-z)\cos(\frac{2{\pi}n}{N})\right]^{1/2},
\end{eqnarray}
where $z_{2}\equiv{\tanh K_{2}}$ and $z\equiv{\alpha}^{2}={\tanh^{2}h}$.
In the case of odd states it is easy to show that the sum $Z^{-}_{1}(h)$,
is equal to $Z^{-}_{1}(h)=2Z^{+}_{1}(h)$.
Finally, we obtain in the thermodynamic limit again the formula
$(1.20)$, $(M\rightarrow N,\;\;\; K_{1}\rightarrow K_{2})$ for free energy on
one spin. 

The $1D$ Ising model was discassed here with so many details because it was 
unexpectedly found, that $Z^+_1(h)$ such as represented in $(1.28)$ can be 
applied in graph theory. Namely, using the representation $(1.28)$ one can 
calculate the generating function for the Hamilton cycles on the simple 
square lattice $(N\times M)$, $\cite{koch95}$. 

\subsection{The $2D$ Ising model}

For further purposes it is convenient to represent the matrix $UT_2$,  
entering the formula $(1.18)$ and where $U$ is defined by $(1.14)$
 in the form of a simple product of matrices $P_n$ such that their
 diagonalization is relatively easy. The reason is that, as is known from
 $\cite{baxter82}-\cite{sml64}$,
 to calculate free energy on one spin in the thermodynamic limit it is
 sufficient to find the maximal eigenvalue of the matrix  $UT_{2}$, which is
  $2^{N}\times 2^{N}$ dimensional.
First of all we note that the matrix $U$ $(1.14)$
 can be represented  in the form of a simple product of matrices
 $U_{0}$:
\begin{equation}
U=\prod^{N}_{n=1}U_{n}=U_{0}\otimes U_{0}\otimes ... \otimes U_{0},\;\;\;\;
N-{\rm factors}, 
\end{equation}
where the matrix $U_{0}$
is defined by the formula $(1.15)$,
in which one should skip the index $n$.
In order  to represent the matrix
$T_{2}$ $(1.6)$
in the form of a simple product we will use the  well known identity
$\cite{glimm81,koch94}$:
\begin{equation}
\exp({\bf A}^{2})=\frac{1}{\pi^{1/2}}\int^{\infty}_{-\infty}\exp(-\xi^{2}
+2{\bf A}\xi)d\xi,
\end{equation}
where ${\bf A}$ --- is a bounded operator (matrix). Writing $\exp(K_{2}
\sigma^{z}_{n}\sigma^{z}_{n+1})$ in the form
\begin{eqnarray*}
\exp(K_{2}\sigma^{z}_{n}\sigma^{z}_{n+1})=\exp\left[\frac{K_2}{2}
(\sigma^{z}_{n}+\sigma^{z}_{n+1})^{2}-K_{2}\right],
\end{eqnarray*}
we can represent the matrix $T_2$ in the form:
\begin{equation}
T_{2}=\frac{e^{-NK_{2}}}{{\pi}^{N/2}}\int^{\infty}_{-\infty} .... 
\int^{\infty}_{-\infty}\exp\left[-\sum^{N}_{n=1}\xi^{2}_{n}+(2K_{2})^{1/2}
\sum^{N}_{n=1}(\xi_{n}+\xi_{n+1})\sigma^{z}_{n+1}\right]\prod^{N}_{n=1}
d\xi_{n},
\end{equation}
where $\sigma^{z}_{N+1}=\sigma^{z}_{1}$ and $\xi_{N+1}=\xi_{1}$.
After writing the matrix  $T_2$, $(1.31)$
 this way we can represent it in the form of a simple product of matrices
 $\exp[(2K_{2})^{1/2}(\xi_{n}+\xi_{n+1})\sigma^{z}]$
inside the integral:
\begin{equation}
\prod^{N}_{n=1}\exp[(2K_{2})^{1/2}(\xi_{n}+\xi_{n+1})\sigma^{z}_{n+1}]=
\prod^{N}_{n=1}\otimes\exp[(2K_{2})^{1/2}(\xi_{n}+\xi_{n+1})\sigma^{z}],
\end{equation}
where
on the right hand side of the formula there is a simple product of $2\times2$
matrices.
Next, we can write the matrix  $UT_{2}$,
using $(1.29)$ and $(1.31-32)$, in the form:
\begin{equation}
UT_{2}=\frac{e^{-NK_{2}}}{{\pi}^{N/2}}\int^{\infty}_{-\infty} .... 
\int^{\infty}_{-\infty}\prod^{N}_{n=1}d\xi_{n}\exp\left[-\sum^{N}_{n=1}
\xi^{2}_{n}\right]\left[\prod^{N}_{n=1}\otimes\exp[(2K_{2})^{1/2}(\xi_{n}+
\xi_{n+1})\sigma^{z}]\right],
\end{equation}
where we included the constant  matrix $U$
under the integral and we used the known theorem on simple product of matrices:
\begin{eqnarray*}
({\bf A_1}\otimes{\bf A_2}\otimes ... )({\bf B_1}\otimes{\bf B_2}\otimes ... )
=({\bf A_{1}B_{1}})\otimes({\bf A_{2}B_{2}})\otimes ...  .
\end{eqnarray*}
Using the expression $(1.33)$ enables calculation of all $2^{ N}$
eigenvalues of the matrix
 $UT_{2}$. Eigenvalues of the matrix $U_{0}\exp[\alpha(\xi_{n}+
\xi_{n+1})\sigma^{z}]$ can be easily calculated and are equal to:
\begin{equation}
\lambda^{\pm}(n,n+1)=e^{\pm\omega(n,n+1)},
\end{equation}
where $\omega(n,n+1)$ is defined as a positive root of the equation:
\begin{equation}
\cosh[\omega(n,n+1)]=\cosh(K^{*}_{1})\cosh[h+\alpha(\xi_{n}+\xi_{n+1})], \;\;
\;\; \alpha\equiv(2K_{2})^{1/2} .
\end{equation}
In the diagonal representation the matrix
 $V$ under the  integral $(1.33)$
can be represented in the form:
\begin{equation}
V=\left[\prod^{N}_{n=1}\otimes S(n,n+1)\right]\prod^{N}_{n=1}\otimes
\left(
\begin{array}{cc}
\lambda^{+}(n,n+1) & 0\\ 0 & \lambda^{-}(n,n+1)
\end{array}\right)
\left[\prod^{N}_{n=1}\otimes S'(n,n+1)\right],
\end{equation}
where $S(n,n+1)S'(n,n+1)={\bf 1}$, and $\lambda^{\pm}(n,n+1)$
are defined above by
$(1.34)$.
From this it follows that the eigenvalues
 $\Lambda_{j}$ of the matrix
$V$ are equal to:
\begin{equation}
\Lambda_{j}=\lambda^{\pm}(1,2)\lambda^{\pm}(3,4) ... \lambda^{\pm}(N,1), \;\;
\;\;\; (j=1,2,3, ...,2^{N}),
\end{equation}
where to each $j$ there corresponds a combination of  $(+)$ and $(-)$
eigenvalues $\lambda^{\pm}(n,n+1)$.

Finally we can express the statistical sum
 $(1.18)$ by the formula:
\begin{equation}
Z(h)=Tr\left[(2\sinh 2K_{1})^{N/2}\frac{e^{-NK_{2}}}{{\pi}^{N/2}}\int^{\infty}_{-\infty} .... 
\int^{\infty}_{-\infty}\prod^{N}_{n=1}d\xi_{n}\exp\left(-\sum^{N}_{n=1}
\xi^{2}_{n}\right)V\right]^{M},
\end{equation}
where the matrix $V$ is given by $(1.36)$.

\section{THE FREE ENERGY}
As we mentioned above, the free energy on  one spin in the thermodynamic limit
can be expressed by the maximal eigenvalue of the matrix
$UT_{2}$, entering $(1.18)$.
We menaged to express this matrix in the form of an $N$-type integral $(1.38)$,
 where the matrix $V$ is defined by
 $(1.36)$, and all the matrix elements of   $V$ are positive.
 On the other hand, in accordance with the known
Frobenius-Perron theorem, the matrix $B$,
with all matrix elements positive has its maximal eigenvalue nondegenerate.
Let us assign to the maximal eigenvalue of the matrix $V$ a letter
$\Lambda_{max}$. In accordance with our definition of the eigenvalues
$\lambda^{\pm}(n,n+1)$, using $(1.37)$ we obtain
the following expression for
$\Lambda_{max}$:
\begin{equation}
\Lambda_{max}=\prod^{N}_{n=1}\lambda^{+}(n,n+1)=\prod^{N}_{n=1}
e^{\omega(n,n+1)}=\exp\left[\sum^{N}_{n=1}\omega(n,n+1)\right],
\end{equation}
where $\omega(n,n+1)$
is defined as a positive root of the equation
$(1.35)$, and
\begin{eqnarray*}
\Lambda_{max}>\Lambda_{j}, \;\;\;\; (j=1,2, ..., 2^{N}).
\end{eqnarray*}
Further we denote the eigenvalues of the  matrix $UT_{2}$ by
${\tilde{\Lambda}}_{j}$. Taking into account the dimension of the matrix
 $UT_{2}$, which is equal to
$2^{N}$, we can write on the base of the relation $(1.18)$
obvious inequalities:
\begin{equation}
{\tilde{\Lambda}}^{M}_{max}\leq Z(h)\leq 2^{N}{\tilde{\Lambda}}^{M}_{max},
\end{equation}
where ${\tilde{\Lambda}}_{max}$ ---
is the maximal eigenvalue in the set ${\tilde{\Lambda}}_j$,
to which we included also the constant factor
 $(2\sinh 2K_{1})^{ N/2}$.
Taking the logarithm $(2.2)$
of this expression and dividing by the nodes number
$NM$, we arrive at the next system of inequalities:
\begin{equation}
\frac{1}{N}\ln({\tilde{\Lambda}}_{max})\leq\frac{1}{NM}\ln Z(h)\leq
\frac{1}{N}\ln({\tilde{\Lambda}}_{max})+\frac{1}{M}\ln 2,
\end{equation}
in which the expression in the middle representations free energy on the node if
we neglect the factor  $-\beta^{-1}$, where $\beta=\frac{1}{k_\beta T}$,
$T$ is temperature.
Going to the limit
$(N,M)\rightarrow\infty$,
we obtain the desired formula describing free energy on one spin in the
thermodynamic limit:
\begin{equation}
f_{2}(h)=-\frac{1}{\beta}\lim_{N,M\rightarrow\infty}\frac{1}{NM}\ln Z(h)=
-\frac{1}{\beta}\lim_{N\rightarrow\infty}\frac{1}{N}
\ln({\tilde{\Lambda}}_{max}),
\end{equation}
where ${\tilde{\Lambda}}_{max}$ is, in accordance with $(1.38)$ and $(2.1)$
equal to:
\begin{equation}
{\tilde{\Lambda}}_{max}=(2\sinh 2K_{1})^{N/2}\frac{e^{-NK_{2}}}{{\pi}^{N/2}}\int^{\infty}_{-\infty} .... 
\int^{\infty}_{-\infty}\prod^{N}_{n=1}d\xi_{n}\exp\left[\sum^{N}_{n=1}
(-\xi^{2}_{n}+\omega(n,n+1))\right].
\end{equation}
Finally, using the Onsager identity:
\begin{equation}
\mid x\mid =\frac{1}{\pi}\int^{\pi}_{0}dq\ln[2\cosh(x)-2\cos(q)],
\end{equation}
we obtain the following expression for the function $\omega(n,n+1)$ $(1.35)$
\begin{equation}
\omega(n,n+1)=\frac{1}{\pi}\int^{\pi}_{0}dq\ln[2\cosh K^{*}_{1}\cosh(h+
(2K_{2})^{1/2}(\xi_{n}+\xi_{n+1}))-2\cos(q)],
\end{equation}
Expressions   $(2.4)$ and $(2.5)$
should describe properly at least the transition to the onedimensional Ising
model. It is easy to  show that the transition to the limit  $K_{2}=0$,
gives the correct result
$(1.20)$ for the onedimensional
Ising model. The analogous limit taken with
respect to the constant
 $K_{1}$ seems a little bit more complicated and has the form:
 \begin{equation}
\lim_{K_{1}\rightarrow 0}{\tilde{\Lambda}}_{max}(K_{1})=
\frac{e^{-NK_{2}}}{{\pi}^{N/2}}\int^{\infty}_{-\infty} .... 
\int^{\infty}_{-\infty}\prod^{N}_{n=1}d\xi_{n}\exp\left[-\sum^{N}_{n=1}
\xi^{2}_{n}\right]\prod^{N}_{n=1}2\cosh[h+\alpha (\xi_{n}+\xi_{n+1})],
\end{equation}
where $\alpha$ is defined above by $(1.35)$.
This is an integral of the Gauss type and it could be relatively easily
calculated. For this purpose we apply the following formal procedure.
Namely, let us write the expression  $2\cosh(...)$,
entering the integral $(2.8)$, in the form:
\begin{equation}
2\cosh[h+\alpha (\xi_{n}+\xi_{n+1})]=\sum_{\mu_{n}=\pm 1}\exp\left[\mu_{n}h+
\alpha\mu_{n}(\xi_{n}+\xi_{n+1})\right],\;\;\; (n=1,2,...,N),
\end{equation}
where we introduced a new variable  $\mu_{n}$
of the Ising type. Therefore, we can represent the right hand side of the
equality $(2.8)$ in the form: 
\begin{eqnarray}
\sum_{(\mu_{n}=\pm 1)}\left\{\frac{e^{-NK_{2}}}{{\pi}^{N/2}}\int^{\infty}_{-\infty} .... 
\int^{\infty}_{-\infty}\prod^{N}_{n=1}d\xi_{n}\exp\left[-\sum^{N}_{n=1}
\xi^{2}_{n}\right]\prod^{N}_{n=1}e^{h\mu_{n}}\exp[\alpha\mu_{n}(\xi_{n}+
\xi_{n+1})]\right\}=\nonumber\\
\sum_{(\mu_{n}=\pm 1)}\exp\left[\sum^{N}_{n=1}(h\mu_{n}+K_{2}\mu_{n}\mu_{n+1})
\right],
\end{eqnarray}
where we took an integral over the variables  $\xi_{n}$, and we imposed on
variables $\mu_{n}$
cyclic boundary conditions $(\mu_{N+1}=\mu_{1})$.
Calculation by standard methods  $\cite{izyum87,thompson88}$ of the sum $(2.10)$,
 and following substitution of the expression
 $(2.4)$, gives  well known result $(1.20)$.

 Consideration of the expressions
 $(2.4)$ and $(2.5)$ for free energy of the twodimensional Ising model in
 external field we present in the end of this paper but now we go to the
 threedimensional case. 

\section{THE THREE-DIMENSIONAL ISING MODEL}

Hamiltonian for the threedimensional Ising model in  external field with 
nearest neighbors interaction we write in the form:
\begin{equation}
{\cal H}=-\sum^{NMK}_{(n,m,k)=1}\left(J_{1}\sigma_{nmk}\sigma_{n,m+1,k}+
J_{2}\sigma_{nmk}\sigma_{n+1,mk}+J_{3}\sigma_{nmk}\sigma_{nm,k+1}+
H\sigma_{nmk}\right),
\end{equation}
where the collective index $(nmk)$
numbers nodes of the simple cubic  lattice and $H$ is the external field.
Constants $I_j$ take into account anisotropy of interaction of Ising spins. We
impose on the variables
$\sigma_{nmk}$, as  it is commonly done, periodic  boundary conditions.
Quantities $N,M$ and $K$ are node numbers in corresponding directions of
a cubic lattice. As is known $\cite{baxter82}$,
the statistical  sum for the threedimensional Ising model can be represented
in the form of a trace of the  $K$-th power  of the fiber-fiber transfer
matrix  $(R)$:
\begin{equation}
W(h)=Tr(R)^{K}\equiv Tr(T_{3}T_{2}T_{1}T_{h})^{K},
\end{equation}
where the matrices $T_{i}$, $(i=1,2,3,h)$ of dimensions $2^{ NM}\times
2^{NM}$ are of the  form:
\begin{equation}
T_{1}=\exp\left(K_{1}\sum_{nm}\sigma^{z}_{nm}\sigma^{z}_{n,m+1}\right), \;\;
\;\; T_{2}=\exp\left(K_{2}\sum_{nm}\sigma^{z}_{nm}\sigma^{z}_{n+1,m}\right),
\end{equation}
\begin{equation}
T_{3}=(2\sinh 2K_{3})^{NM/2}\exp\left(K^{*}_{3}\sum_{nm}\sigma^{x}_{nm}\right),
\;\;\;\; T_{h}=\exp\left(h\sum_{nm}\sigma^{z}_{nm}\right).
\end{equation}
Here $K_{i}=\beta J_{i}$, $(i=1,2,3)$; $\beta =(1/k_{B}T)$, $T$ -
temperature, $h=\beta H$, and $K_{3}$ and $K^{*}_{3}$
are connected by relations of type $(1.8)$.
In  the formulaes $(3.3--4)$ the matrices $\sigma^{x,z}_{nm}$ are Pauli  
matrices, which are defined  analogously to
$(1.9)$, and have dimensions $2^{NM}\times 2^{NM}$.

Continuing considerations analogous to these in the  twodimensional case we
obtain the following  formula describing  free energy  on one spin  in the
thermodynamic limit:
                     
\begin{equation}
f_{3}(h)=-\frac{1}{\beta}\lim_{N,M\rightarrow\infty}\frac{1}{NM}\ln\Lambda_{max},
\end{equation}
where the maximal eigenvalue $\Lambda_{max}$ of the matrix $R$, $(3.2)$
is defined by:
\begin{eqnarray*}
\Lambda_{max}=
\end{eqnarray*}
\begin{equation}
(2\sinh 2K_{3})^{NM/2}\frac{e^{-NM(K_{1}+K_{2})}}{\pi^{NM}}
\int^{\infty}_{-\infty}...\int^{\infty}_{-\infty}\prod_{nm}d\eta_{nm}d\xi_{nm}
\exp\left[\sum^{NM}_{n,m=1}(-\eta^{2}_{nm}-\xi^{2}_{nm}+\omega(n,m))\right],
\end{equation}
and $\omega(n,m)$ is defined aa the positive root of the equation:
\begin{equation}
\cosh\omega(n,m)=\cosh K^{*}_{3}\cosh[h+\alpha_{1}(\eta_{n+1,m}+\eta_{n+1,m+1})
+\alpha_{2}(\xi_{n,m+1}+\xi_{n+1,m+1})],
\end{equation}
where $\alpha_{1,2}=(2K_{1,2})^{1/2}$.

We impose on integration variables
 $\eta_{nm}$ and $\xi_{nm}$
 cyclic boundary conditions, in accordance with periodic boundary conditions for
 the former variables:
 \begin{equation}
\sigma^{z}_{N+1,m}=+\sigma^{z}_{1,m}, \;\;\;\;\sigma^{z}_{n,M+1}=
+\sigma^{z}_{n,1}, \;\;\; n(m)=1,2,3,....N(M).
\end{equation}
Similarly  as in the twodimensional case, the function
 $\omega(n,m)$
can be expressed  explicitly  in terms of variables  $\eta_{nm}$ and $\xi_{nm}$,
 using for this aim the integral representation given by the Onsager identity
 $(2.6)$.

It is a little bit more complicated  matter to take  a limit  with respect to
the constant of interaction $K_3$, although the derived formula  is much 
simpler than  the  formulae  $(2.4-5)$,
 Imposition of the limit
 $(K_{3}\rightarrow 0)$ in the formulae $(3.5-7)$,
  gives, after  some simple transformations, the following  representation of
  free energy on one spin of the twodimensional Ising model
  $f_{2}(h)$:
\begin{eqnarray*}
f_{2}(h)=-\frac{1}{\beta}(\lim_{K_{3}\rightarrow 0, (N,M)\rightarrow\infty})
\frac{1}{NM}\ln\Lambda_{max}=
\end{eqnarray*}
\begin{eqnarray*}
-\frac{1}{\beta}\lim_{(N,M)\rightarrow\infty}\frac{1}{NM}\ln\left\{
\frac{e^{-NM(K_{1}+K_{2})}}{\pi^{NM}}
\int^{\infty}_{-\infty}...\int^{\infty}_{-\infty}\prod_{nm}d\eta_{nm}d\xi_{nm}
\exp[-\sum_{nm}(\eta^{2}_{nm}+\xi^{2}_{nm})]\right.
\end{eqnarray*}
\begin{equation}
\left.\times\prod_{nm}2\cosh[h+\alpha_{1}(\eta_{n+1,m}+\eta_{n+1,m+1})
+\alpha_{2}(\xi_{n,m+1}+\xi_{n+1,m+1})]\right\},
\end{equation}
where we used relations of the type $(1.8)$, and $\alpha_{1,2}$
are defined above $(3.7)$. The integrals in $(3.9)$ are integrals
of the Gauss type and, as it is easy to show applying the
described above formal way  of introducing  a variable  of Ising type
$\mu_{nm}=\pm 1$, can be represented in the form $(1.2)$.

Analogously, one can show rigorously that free energy on one spin  for the
threedimensional Ising model can be  represented  in the form of a multiple
integral  of the Gauss type:
 \begin{eqnarray*}
f_{3}(h)=-\frac{1}{\beta}\lim_{(N,M,K)\rightarrow\infty}\frac{1}{NMK}\ln
\left\{
\frac{e^{-NMK(K_{1}+K_{2}+K_{3})}}{\pi^{3NMK/2}}\int^{\infty}_{-\infty}...
\int^{\infty}_{-\infty}\prod_{nmk}d\eta_{nmk}d\xi_{nmk}d\zeta_{nmk}\right.
\end{eqnarray*}
\begin{eqnarray*}
\left.\times\exp[-\sum_{n,m,k}(\eta^{2}_{nmk}+\xi^{2}_{nmk}+\zeta^{2}_{nmk})]
\right.
\end{eqnarray*}
\begin{equation}
\left.\times\prod_{nmk}2\cosh[h+\alpha_{1}(\eta_{nmk}+\eta_{n+1,mk})
+\alpha_{2}(\xi_{nmk}+\xi_{n,m+1,k})+\alpha_{3}(\zeta_{nmk}+\zeta_{nm,k+1})]
\right\},
\end{equation}
where $\alpha_{i}=(2K_{i})^{1/2}$, $(i=1,2,3)$.
The formulae $(3.10)$  can be in obvious way generalized to describe 
d-dimensional Ising  models but we  will not stop  on these matters here. 

We will make here a few  remarks. First of
all, as far as it is known  to the author, the representations for free energy
on one spin in the forms  $(2.4-5)$ and
$(3.5-10)$ for
the twodimensional and threedimensional Ising models, respectively, did not
appear in the literature. We belive that the known representations  (see,
e.g. $\cite{mccoy-wu73,barouch80,wilson74}$)
are more complicated than the ones derived by us. For example, in the paper
$\cite{wilson74}$ was given the following integral representation for the 
Ising model:
\begin{equation}
Z=\prod_{\bf m}\int^{\infty}_{-\infty}\exp(-\frac{1}{2}bs^{2}_{\bf m}-
us^{4}_{\bf m})\exp(K\sum_{\bf n,i}s_{\bf n}s_{\bf n+i})ds_{\bf m},
\end{equation}
where $(\bf m,n,i)$ ---- are vector indices
( we use notation from the paper
$\cite{wilson74}$). In the limit  $u\rightarrow\infty$ and
$b\rightarrow{-\infty}$, $(b=-4u)$
the Ising model is recovered (for this aim one should use  additionally the fact
that to every particular spin there is a factor 
$(u/\pi)^{1/2}\exp(-u)$).
 It is clear that the presence of the term
 $us^{4}_{\bf m}$ in
$\exp(...)$ of the expression $(3.11)$
considerably complicates analysis of this expression. Further,  although the
formulae  $(3.9)$ and $(3.10)$
are in  a sense obvious, the formulae                         
$(2.4-5)$ and $(3.5-7)$
 are not obvious. The integrals
 $(2.5)$ and $(3.6)$
can be represented as  integrals of the "quasi Gauss" type,  because the functions
 $\omega(n,n+1)$ and
$\omega(n,m)$,
described by the relations   $(1.35)$
and $(3.7)$,
respectively, in accordance with the Onsager identity
 $(2.6)$
 are almost "linear" in their arguments
 $(\xi_{n})$ and $(\eta_{nm},\xi_{nm})$.
This justifies  our hopes we could learn now to calculate rigorously the
integrals
$(2.5)$ and $(3.6)$ in case $h=0$
using an Ising type variable
$(\mu=\pm 1)$,
described above. On the other hand, it seems to us that for the case
 $(h\neq0)$ it is much simpler to deal with the expression  $(2.5)$,
 than with the expression  $(3.9)$,
although it could sound paradoxically. For the threedimensional Ising model in
external field  $(h\neq0)$
the situation is no longer  so clear, for the infinitely small field
$(h\sim 0)$, similarly as in the twodimensional case, it is easier to analyze
the expression $(3.6)$, than $(3.10)$, that we shall be show in a next 
publication. 

\section{CONCLUSIONS}

The derived expressions   $(2.4-5)$ and $(3.5)-(3.10)$
for free energy on one spin  for the Ising  model can be of some interest, we
hope. The reason is we actually should learn  how to calculate logarithmic
asymptotics of multiple integrals of Gauss type, as it could be seen from
 $(2.4)$ and $(3.5)$. It is known, that there exist a well developed formalism
 of calculation of logarithmic asymptotes for integrals  of the Laplace type 
for the onedimensional as well as for the multidimensional cases
$\cite{fedoryuk87}$.
In the case under consideration the situation is a little bit more 
complicated, because for the integrals of the kinds  $(2.4)$ and
$(3.5)$  it is not possible to transform them to a form of a multiple integral
of the Laplace type at least  in the framework  of their classical definition.
With the increase of the large parameter
$(\lambda\rightarrow\infty)$
there changes also the number of variables, over which one integrates and this
needs reformulation of the corresponding  methods of asymptotic estimation of
the considered integrals
$\cite{fedoryuk87}$.
In future publications we intended to investigate in more details the
expressions  $(2.4)$ and
$(3.5)$, obtained in this paper and to calculate critical  indices for the 
Ising model.

\acknowledgements

I am grateful to H. Makaruk and R. Owczarek for help in preparation of the final
form of this paper.

This paper was supported by the KBN grant $ N^o$ {\bf 2 P03B 088 15}.

\end{document}